\documentstyle[prl,aps,twocolumn,epsf]{revtex}
\def\break#1{\pagebreak \vspace*{#1}}
\begin{document}

\draft

\title{Microtubules: Montroll's kink and Morse vibrations}

\author{H.C. Rosu\cite{byline}
}

\address{
{Instituto de F\'{\i}sica de la Universidad de Guanajuato, Apdo Postal
E-143, Le\'on, Gto, M\'exico}\\
{Institute of Gravitation and Space Sciences, P.O. Box MG-6,
Magurele-Bucharest, Romania}
}

\maketitle
\widetext

\begin{abstract}

Using a version of Witten's supersymmetric quantum mechanics
proposed by Caticha, we relate Montroll's kink to a traveling,
asymmetric Morse double-well potential suggesting in this way
a connection between kink modes and vibrational degrees of freedom
along microtubules.

\end{abstract}

{\scriptsize \pacs{PACS \# 's:
87.15.He, 03.65.Ge, 11.30.Pb \hspace{0.7cm}  LAA \# : cond-mat/9606109
\hspace{0.7cm} Phys. Rev. E 55 (Febr. 1997) 2038-39}}


\narrowtext


Two decades ago, Collins, Blumen, Currie and Ross \cite{1} studied the
dynamics of domain walls
in ferrodistortive materials. They started with a Ginzburg-Landau
hamiltonian/free-energy with driven field and dissipation included leading
to the Euler-Lagrange dimensionless equation of motion
\begin{equation} \label{1}
\psi ^{''}+\rho\psi ^{'}-\psi ^3 +\psi+\sigma=0~,
\end{equation}
where the primes are derivatives with respect to a traveling coordinate
$\xi =x-v_Kt$, $\rho$ is a friction coefficient and $\sigma$ is related to
the driven field \cite{1}.
This equation is a traveling coordinate transform of a nonlinear
reaction-diffusion equation widely studied
in the contexts of population genetics and
nonequilibrium chemical systems. It has also drawn the attention of
pure mathematicians who proved many general results.

Montroll \cite{mont} showed that
Eq.~(1) has a unique bounded (kink-like) solution as follows
\begin{equation}  \label{2}
\psi(\xi)=a+\frac{b-a}{1+\exp(\beta\xi)}~,
\end{equation}
where  $\beta=(b-a)/\sqrt{2}$
and the parameters $a$ and $b$ are two of the solutions of the cubic equation
\begin{equation} \label{3}
(\psi -a)(\psi -b)(\psi -d)=\psi ^3 -\psi -\sigma~.
\end{equation}
Recently, Montroll's kink has been used as an energy-transfer mechanism
in microtubules \cite{mtub1,mtub2} and it is to this particularly interesting
biological context that we would like to apply our remarks in the
following.
But firstly,
it is essential to notice that
Montroll's kink can be written as follows
\begin{equation} \label{4}
\psi(\xi)=\frac{\beta}{\sqrt{2}}\Bigg[\left(1+\frac{a\sqrt{2}}{\beta}\right)-
\tanh\left(\frac{\beta\xi}{2}\right)\Bigg]
\end{equation}
and an obvious rescaling leads to the useful form
\begin{equation} \label{5}
K(\xi)=
\gamma -\tanh\left(\frac{\beta \xi}{2}\right)~,
\end{equation}
\break{1.197in}
where $\gamma =1+\frac{a\sqrt{2}}{\beta}$.
Eq.~(5) is a requisite in order to enter a construction method of exactly
soluble
double-well potentials in the Schr\"odinger equation proposed by Caticha
\cite{cat}.
The scheme is a non-standard application of
Witten's supersymmetric quantum mechanics \cite{w}
having as the essential assumption the idea of considering the kink as
the switching function between the two lowest eigenstates of the
Schr\"odinger equation with a double-well potential. Thus
\begin{equation} \label{6}
\phi _1=K\phi _0~,
\end{equation}
where $\phi _{0,1}$ are solutions of $\phi ^{''}_{0,1}+[\epsilon _{0,1}-u(\xi)]
\phi _{0,1}(\xi)=0$, and $u(\xi)$ is the double-well potential to be found.
Substituting the assumption Eq.~(6) into the Schr\"odinger equation for
the subscript 1 and substracting the same equation multiplied by the
switching function for the subscript 0, one obtains
\begin{equation} \label{7}
\phi ^{'}_{0}+W\phi _0=0~,
\end{equation}
which is the basic equation introducing the superpotential $W$ in
Witten's supersymmetric quantum mechanics.
In the present approach $W$ is given by
\begin{equation} \label{8}
W=\frac{K^{''}+\epsilon K}{2K^{'}}~,
\end{equation}
where $\epsilon=\epsilon _1-\epsilon _0$ is the lowest energy splitting in
the double-well Schr\"odinger equation. The ground-state wave function is
of the supersymmetric type
\begin{equation} \label{9}
\phi _0(\xi)=\phi _0(0)\exp\Bigg[-\int_0^{\xi} W(y)dy\Bigg]~,
\end{equation}
where $\phi _0(0)$ is a normalization constant. The double-well potential
is determined up to an additive constant by the `bosonic' Riccati equation
\begin{equation} \label{10}
u(\xi)=W^2-W'+\epsilon _0~,
\end{equation}
which is another basic supersymmetric formula.
For Montroll's kink the superpotential will be
\begin{equation} \label{11}
W(\xi)=-\beta\tanh(\beta\xi)+\frac{\epsilon}{4\beta}\Bigg[\sinh(2\beta\xi)+
2\gamma\cosh ^2(\beta\xi)\Bigg]
\end{equation}
and the ground-state Schr\"odinger function reads
\begin{eqnarray} \nonumber
\phi _0&(\xi)& =\phi _0(0)\cosh(\beta\xi) \\
        &\times&\exp\Bigg[-\frac{\epsilon}{4\beta ^2}
\sinh ^2(\beta \xi)+\gamma \beta\xi +\frac{\gamma}{2}\sinh(2\beta\xi)\Bigg]~,
\end{eqnarray}
while $\phi _1$ is obtained by switching the ground-state wave function with
the kink. If, as suggested by Caticha, one chooses the ground state energy
to be
\begin{equation} \label{13}
\epsilon _0=-\beta ^2-\frac{\epsilon}{2}+\frac{\epsilon ^2}{32\beta ^2}
\left(\gamma ^2-1\right)~,
\end{equation}
then $u(\xi)$ is a traveling, asymmetric Morse double-well potential of
depths
\begin{equation} \label{14}
U_0^{L,R}=4\beta ^2\Bigg[1\pm \frac{2\epsilon \gamma}{(4\beta)^2}\Bigg]~,
\end{equation}
where the superscripts stand for left and right well.
The difference in depth, the bias, is
$\Delta\equiv U_0^L-U_0^R=2\epsilon\gamma$, while the location of the
potential minima on the traveling axis is at
\begin{equation}  \label{15}
\xi _{L,R}=\mp\frac{1}{2\beta}\ln \Bigg[\frac{(4\beta)^2\pm
2\epsilon\gamma}{\epsilon(\gamma\mp 1)}\Bigg]~.
\end{equation}

Thus, there are two interpretations of the kink in Eq.~(5), either as a
propagating domain wall through a sequence of on-site quartic
double-well potentials, or as a kink connected to a propagating
vibrational Morse double-well. The latter picture is closer to the spirit
of bioenergetics if one remembers Davydov model where both the vibrational
soliton and the phonon kink are of course propagating objects.
The Morse double-well picture, though
quite appealing, occurs only when the set of relationships Eq.~(13-15)
is fulfilled by the parameters of the kink and the Morse parameters and
therefore some experimental evidence is required.
As discussed in \cite{mtub1}, in the static on-site picture,
the mobile electron on each dimer unit may be localized either more toward
the $\alpha$ monomer or more toward the $\beta$ one, and in fact the kink
is just contributing to the electron tunneling between the two states of the
double-well on-site potential. Thus, in the case of the Morse picture,
one may think of molecular photoelectron
spectroscopy of tubulin dimers performed with a simple retarding-field
photoelectron spectrometer, similar for example to that of Price and
Ibrahim \cite{pi}.
Moreover, a Morse double-well parametrization
of the potential curves corresponding to the differential photoelectron
spectrum should be performed. As far as I know, there are no such experimental
data at the present time.
I recall that in elaborating a Fr\"ohlich-like (coherent) model of MTs,
Samsonovich, Scott, and Hameroff \cite{ssh} quoted only two {\em indirect}
experimental evidence for the existence of coherent excitations along MTs,
namely the 2.45 GHz irradiation experiment of
Neubauer {\em et al.} \cite{n} and the microtubule associated protein
attachment site superlattices on MTs  \cite{kim}.

In conclusion, according to Caticha's scheme,
experimental evidence of Morse-type vibrations would support both
a {\em traveling} Morse double-well potential and the existence of a
(traveling) kink in MTs with the parameters of the kink
depending on the Morse parameters. On the other hand, the same kink parameters
may be interpreted in terms of the parameters of a quartic double-well on-site
potential (the original Montroll kink).
In other words, if one of the interpretations fails the kink is
still there, while if both are correct further insights into their connection
must be provided.

The supersymmetric method is quite general and can be applied to other kinks
as well, and as a matter of fact, to any model where a traveling potential
is preferred. The weak point is that the ground state energy should be fixed
rather arbitrarily for each case in terms of the kink parameters $\beta$ and
$\gamma$ and the tunneling splitting $\epsilon$.

\section*{Acknowledgments}
This work was partially supported by the CONACyT Project
4868-E9406. Discussions with J.A. Tuszy\'nski have been useful.


\end{document}